\documentclass[10pt,twocolumn]{aastex631}

\usepackage[utf8]{inputenc}
\usepackage[flushleft]{threeparttable}
\usepackage{longtable}
\usepackage{overpic}
\usepackage{gensymb}
\usepackage{multirow}
\usepackage{array}
\usepackage{amsmath}
\usepackage{mathtools}

\shorttitle{Light curve grid of stripped-envelope supernovae}
\shortauthors{Fang et al.}

\begin{document}

\title{A Radiation-Hydrodynamic Light Curve Grid and Interpolation Framework for Stripped-Envelope Supernovae}
\author[0000-0002-1161-9592]{Qiliang Fang}\affiliation{National Astronomical Observatory of Japan, National Institutes of Natural Sciences, 2-21-1 Osawa, Mitaka, Tokyo 181-8588, Japan}

\author[0000-0003-1169-1954]{Takashi J. Moriya}
\affiliation{National Astronomical Observatory of Japan, National Institutes of Natural Sciences, 2-21-1 Osawa, Mitaka, Tokyo 181-8588, Japan}
\affiliation{Graduate Institute for Advanced Studies, SOKENDAI, 2-21-1 Osawa, Mitaka, Tokyo 181-8588, Japan}
\affiliation{School of Physics and Astronomy, Monash University, Clayton, VIC 3800, Australia}

\begin{abstract}
We present a grid of {7,560} stripped-envelope supernovae (SESNe) light curves based on radiation-hydrodynamic simulations of exploding helium-star progenitors. The grid spans an ejecta-mass range representative of typical SESNe, together with broad ranges of explosion energies, radioactive nickel masses, and degrees of material mixing. We systematically investigate how these physical parameters shape the light curves and develop an interpolation-based parameter inference framework that enables the application of the model grid to observational data. Using the simulated light curves as mock observations, we assess {whether the input parameters can be recovered by} the widely used Arnett model. Although it reproduces the overall light-curve morphology, the inferred ejecta masses show no significant correlation with the input values. Moreover, the analytical model produces a spurious correlation between the inferred explosion energy and nickel mass, despite these parameters being independent in the underlying model grid. These results demonstrate that neglecting physical processes naturally included in radiation-hydrodynamic models can lead to biased parameter inference, even when the overall light-curve morphology is well reproduced, and highlight the importance of physically motivated radiation-hydrodynamic models for the interpretation and the population study of SESNe.

\end{abstract}
 


\section{INTRODUCTION}
Stripped-envelope supernovae (SESNe) are the explosions of massive stars that have lost most or all of their hydrogen-rich envelopes prior to core collapse, as indicated by the absence of hydrogen features in their spectra (\citealt{nomoto95,filippenko97}). These events provide important constraints on the final stages of massive-star evolution and the physics of core-collapse explosions.

The rapid growth of time-domain surveys has dramatically increased the number of well-observed SESNe. Large observational samples now make it possible to investigate the statistical distributions of explosion parameters and their correlations with progenitor properties. Extracting such information, however, requires reliable methods for translating observed light curves into physical quantities.

Analytical models, developed by \citet{arnett82} and later refined by \citet{valenti08}, have been widely used for this purpose. Owing to their computational efficiency, these Arnett-type models have become a standard tool for estimating progenitor and explosion properties from observed light curves. Numerous studies have applied them to large SESN samples to explore population-level trends, including the distributions of physical parameters and correlations among them, thereby providing insight into the nature of SESN progenitors and their explosion mechanisms \citep{drout11,lyman16,taddia18,prentice19,barbarino21,karamehmetoglu23,zhao26}. However, the analytical framework relies on several simplifying assumptions, including constant opacity, centrally concentrated radioactive heating, and a one-zone description of the ejecta. Consequently, the extent to which the inferred parameters reflect the true physical properties, and whether systematic biases are introduced, remains important topics of investigation.

An alternative approach is to compare observations directly with numerical radiation-hydrodynamic models. Such calculations incorporate a more realistic treatment of the ejecta structure, radioactive energy deposition, recombination, and radiative-transfer effects, thereby providing a physically self-consistent description of the observed emission \citep{bersten12,taddia18,lu25,yadavalli26}. Nevertheless, generating sufficiently large model grids for parameter inference remains computationally expensive, and the discrete nature of available simulations complicates their direct application to observational data.

To address these challenges, we construct a large grid of radiation-hydrodynamic light-curve models of SESNe, based on the explosions of helium-star progenitors. The grid spans a range of ejecta masses motivated by constraints from late-phase spectra, together with broad ranges of explosion energies and radioactive nickel masses, encompassing much of the diversity observed among SESNe. Rather than restricting parameter estimation to the discrete model grid, we develop an interpolation framework that enables continuous exploration of the parameter space. This approach combines the physical realism of numerical models with the computational efficiency required for fitting observational data.

The interpolation framework is then applied to two well-observed SNe Ib, SNe 2007Y and 2009jf, to infer their physical properties through a Markov chain Monte Carlo (MCMC) analysis. This demonstrates the practical application of the model grid for the physical interpretation of observed SESNe light curves.

Finally, the model grid is used as a controlled laboratory to evaluate the performance of Arnett-type analytical light-curve models. The numerical light curves are treated as mock observations and fitted with the Arnett-type model, which allows for a systematic assessment of the inherent biases and limitations. Because the underlying physical parameters of the numerical models are known by construction, the model grid provides an ideal benchmark for testing the reliability of this widely used analytical framework.

This paper is organized as follows. In \S2, we describe the numerical setup for the progenitor and light-curve calculations. The general properties of the model grid are presented in \S3. In \S4, we develop the interpolation framework and perform a series of injection-recovery tests to evaluate its accuracy, which is then applied to observational data. In \S5, we use the Arnett-type models to fit the mock observations generated from the numerical light curves and quantify the systematic uncertainties and biases of the analytical approach. Our conclusions are summarized in \S6.

\section{Numerical Setup}
In this section, we briefly introduce the numerical setup of the progenitor, explosion and the radiation-hydrodynamic models.

\subsection{Progenitor Evolution}
We use the one-dimensional stellar evolution code, Modules for Experiments in Stellar Astrophysics ($\texttt{MESA}$ version r23.05.01; \citealt{paxton11, paxton13, paxton15, paxton18, paxton19}) to compute massive star evolution. The settings are the same as those outlined in \cite{fang25a,fang25b}: non-rotating, solar metallicity progenitor models with zero-age-main-sequence (ZAMS) masses $M_{\rm ZAMS}$ ranging from 11.5, 12.0 to 20.0\,$M_{\rm \odot}$ in 1\,$M_{\rm \odot}$ steps.

The adopted micro physics, such as overshooting parameters, are the same as those adopted in \cite{fang25a,fang25b} so we do not repeat here. With these setups, the progenitor models are evolved from the pre-main-sequence stage to core helium depletion without including wind mass loss. We then use the command \texttt{relax\_mass\_to\_remove\_H\_env} to artificially remove the entire hydrogen-rich envelope. The resulting stripped progenitors (bare helium stars) subsequently evolved further to core oxygen depletion, again without wind mass loss. We do not evolve the models to core collapse, as calculations during the advanced burning stages become computationally expensive, while the innermost products are largely excised once the explosions are phenomenologically triggered (\S~2.2), making them irrelevant for the present study. 

We note that the choice of mixing prescriptions and the adopted Wolf-Rayet wind mass-loss rates (zero in this work; but see \citealt{ertl20,woosley21}) will significantly affect the final helium-core mass for a given $M_{\rm ZAMS}$ (\citealt{temaj24}), the variation in $M_{\rm ZAMS}$ should primarily be regarded as a means of generating models with different ejecta masses ($M_{\rm ej}$). The correspondence between $M_{\rm ej}$ and $M_{\rm ZAMS}$ should not be interpreted literally.

\subsection{Explosion and Radiative-Transfer}
After the hydrostatic evolution of the progenitor models, we use the hydrodynamic mode in $\texttt{MESA}$ to trigger the explosion. 

We first perform the mass-cut to mimic the formation of the compact remnant. The removed mass $M_{\rm cut}$ depends on $M_{\rm ZAMS}$, and is selected based on the inner mass coordinate following the criteria in \citet{fang25a}. For models with $M_{\rm ZAMS}$\,=\,12\,$M_{\rm \odot}$, $M_{\rm cut}$\,$\sim$\,1.5\,$M_{\rm \odot}$ and more massive progenitors will have higher $M_{\rm cut}$. We emphasize that the choice of $M_{\rm cut}$ is somewhat arbitrary and serves primarily as a practical prescription for defining the compact remnant. Consequently, the conversion from the $M_{\rm ej}$ back to the helium core mass remains subject to systematic uncertainty.

After the remnant is removed, the explosion energy is manually put in the inner 0.2\,$M_{\rm \odot}$ as thermal bomb to trigger the explosion. The deposited energy is sampled with non-uniform intervals of 0.2 to 0.5\,$\times\,10^{51}$\,erg, and the range is determined by the progenitor models, such that the ratio between the asymptotic energy (i.e., the energy stored in the expanding ejecta, equal to the deposited energy minus the binding energy) and the ejecta mass $M_{\rm ej}$ spans 0.13-0.83 in units of $10^{51}$\,erg/$M_{\rm \odot}$. Hereafter, for simplicity, we use the term "explosion energy" or $E_{\rm K}$ to refer to the asymptotic energy.

After the launch of the explosion, we track the shock propagation until it reaches a depth of 0.05\,$M_{\rm \odot}$ below the stellar surface, at which point the hydrodynamic simulation is terminated. We then manually insert different amounts of $^{56}$Ni into the ejecta, with $M_{\rm Ni}$ ranging from 0.03 to 0.30\,$M_{\rm \odot}$, sampled with non-uniform intervals of 0.01 to 0.03\,$M_{\rm \odot}$. Following \citet{moriya20}, the degree of material mixing is parameterized by $M_{\rm mix}$, below which all elements, including $^{56}$Ni, are assumed to be uniformly mixed. For each ejecta model, we vary $M_{\rm mix}$ such that 30\% to 80\% (with a step size of 10\%) of the ejecta is mixed in mass coordinate. The fraction of the mixed mass is denoted as $f_{\rm mix}$. The MESA inlists used to generate the progenitor and explosion models are available on Zenodo under an open-source Creative Commons Attribution 4.0 International license: doi:10.5281/zenodo.13953755.

With the above setups, the models are handed off to $\texttt{STELLA}$, a one-dimensional multi-group radiation-hydrodynamics code (\citealt{blinnikov98, blinnikov00, blinnikov06}), for the calculation of the light curves. 

$\texttt{STELLA}$ implicitly couples the hydrodynamics to the time-dependent radiation field over multiple frequency groups. The radiation field itself is not assumed to be in local thermodynamic equilibrium (LTE). The material ionization and excitation states used in calculating the opacities and emissivities, however, are evaluated assuming LTE. The opacity includes bound-free, free-free, electron-scattering, and bound-bound contributions, with the large number of spectral lines treated through the expansion-opacity. Consequently, $\texttt{STELLA}$ is designed primarily for radiation-hydrodynamic evolution and broad-band light-curve calculations. Its predictions are expected to be most reliable during the optically thick and photospheric phases. In this work, we therefore restrict our interpretation of the $\texttt{STELLA}$ calculations to the bolometric light curve during the photospheric phase.

We set 600 spatial zones and 40 frequency bins which span from 1 to 50,000\,$\rm \AA$ uniformly on the logarithm scale. After shock breakout, the velocity of the outermost material can approach the speed of light, $c$. Following \citet{moriya20}, material with velocities exceeding 0.1\,$c$ is removed, therefore the shock-breakout phase is not modeled accurately. However, since this phase is not relevant to the present work, we ignore its effects. The maximum velocity of the ejecta models are therefore restricted to 3$\,\times\,10^4$\,km\,s$^{-1}$, while the minimum velocity is determined by the radiation-hydrodynamics, which typically ranges from $\sim$\,500 to 1000 km\,s$^{-1}$.

\begin{table}[!b]
\begin{center}
\caption{Progenitor models. The numbers in the parentheses represent the numbers of the grid points.}
\label{tab:grid_params}
\begin{tabular}{cccccc}
\hline
$M_{\rm ZAMS}$&$M_{\rm He\,core}$&$M_{\rm ej}$&$E_{\rm K}$&$M_{\rm Ni}$&$f_{\rm mix}$\\
$M_{\rm \odot}$&$M_{\rm \odot}$&$M_{\rm \odot}$&$10^{\rm 51}$\,erg&$M_{\rm \odot}$&\\
\hline
11.5&2.956&1.544&0.3-1.5(6)&&\\
12.0&3.150&1.628&0.2-1.5(7)&&\\
13.0&3.522&1.907&0.3-1.8(7)&&\\
14.0&3.951&2.337&0.3-2.0(8)&&\\
15.0&4.354&2.537&0.3-2.4(10)&0.03-0.3&0.3-0.8\\
16.0&4.770&2.955&0.3-2.8(12)&(12)&(6)\\
17.0&5.225&3.410&0.3-3.0(13)&&\\
18.0&5.687&3.872&0.5-3.6(13)&&\\
19.0&6.148&4.332&0.5-3.6(14)&&\\
20.0&6.613&4.792&0.5-4.0(15)&&\\
\hline%
\end{tabular}
\end{center}
\end{table}

\begin{figure}
\epsscale{1}
\plotone{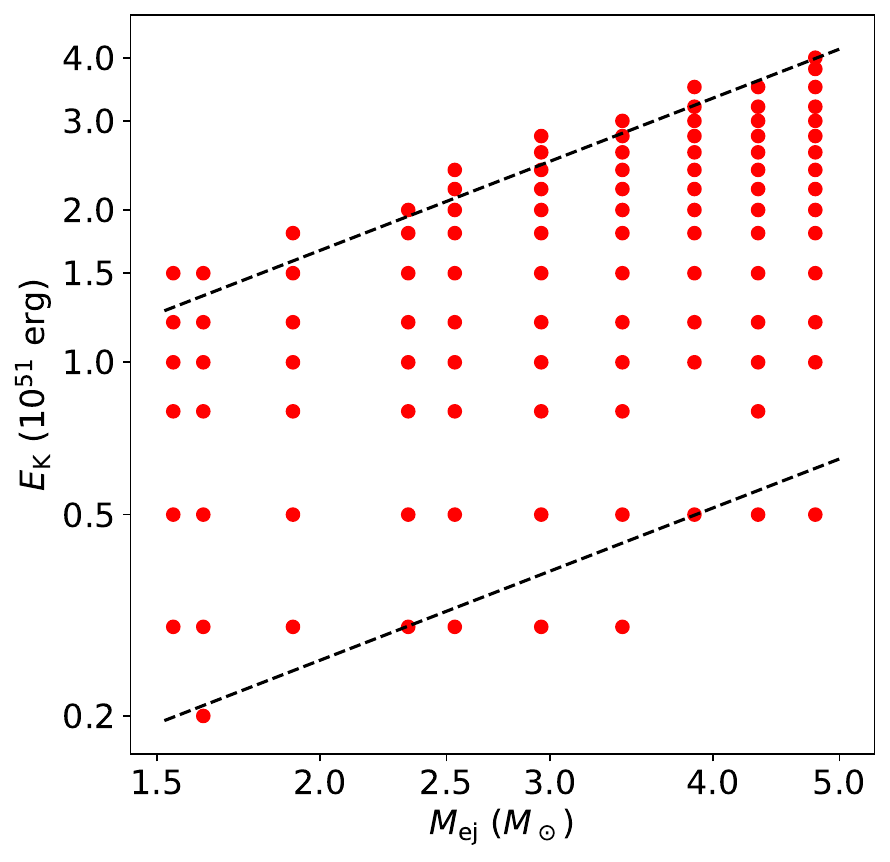}
\centering
\caption{The $M_{\rm ej}-E_{\rm K}$ space of the grid. The black dashed lines corresponds to $E_{\rm K}/M_{\rm ej}$\,=\,0.13 and 0.83 (in units of $10^{51}$\,erg/$M_{\rm \odot}$).}
\label{fig:Mej_Ek_space}
\end{figure}

Table~1 summarizes the properties of the model grid, while the $M_{\rm ej}$-$E_{\rm K}$ parameter space is shown in Figure~\ref{fig:Mej_Ek_space}. In total, the grid contains 7,560 models.

We note that the explored ejecta mass ($M_{\rm ej}$) spans a relatively narrow range (similar to the models in \citealt{dessart15}). This choice is motivated by constraints from late-phase observations. More massive helium stars generally synthesize larger amounts of oxygen during stellar evolution (see, e.g., \citealt{takahashi23}). The oxygen mass in the ejecta can therefore serve as a proxy of the mass of the progenitor helium star and, consequently, the the ejecta mass, apart from variations in the compact-remnant mass and possible fallback. Based on analysis of the relative strength of oxygen emission lines in nebular spectra of large samples of SESNe, \cite{fang19,fang22} found that the majority of SNe IIb/Ib (helium-rich SNe thought to arise from the explosions of helium stars) exhibit weaker oxygen emission than the 17A model of \cite{jerkstrand_15}, which has an ejecta mass of 3.5\,$M_{\odot}$. This comparison suggests that most SNe IIb/Ib likely have ejecta masses below this value. We therefore expect that the adopted $M_{\rm ej}$ range encompasses the typical ejecta masses of helium-rich SESNe.

\section{Model Survey}
In this section, we investigate how the four major parameters, namely the ejecta mass $M_{\rm ej}$, the specific kinetic energy $\eta$\,($\coloneqq E_{\rm K}$/$M_{\rm ej}$), the $^{56}$Ni mass $M_{\rm Ni}$, and the degree of mixing $f_{\rm mix}$, affect the properties of the bolometric light curves. Here we adopt $\eta$ instead of $E_{\rm K}$ as a primary parameter because, as demonstrated by \cite{dessart16}, it is more fundamentally connected to the light-curve properties. Throughout this work, $\eta$ is in the unit of 10$^{51}$\,erg/$M_{\rm \odot}$. 

The bolometric light curves are characterized by four key quantities: the peak luminosity ($L_{\rm peak}$), the peak time ($t_{\rm peak}$), the rise time ($t_{\rm rise}$), and the decline time ($t_{\rm decline}$). Here, $t_{\rm rise}$ is defined as the time before $t_{\rm peak}$ at which the bolometric luminosity is fainter than $L_{\rm peak}$ by 0.4\,dex (1.0 mag), while $t_{\rm decline}$ is defined as the time after $t_{\rm peak}$ at which the bolometric luminosity has declined by 0.4\,dex.

The light curve shape, $S$, is defined on normalized phase:
\[
\tau =
\begin{cases}
\dfrac{t - t_{\rm peak}}{t_{\rm rise}}, & t < t_{\rm peak}, \\[8pt]
\dfrac{t - t_{\rm peak}}{t_{\rm decline}}, & t \geq t_{\rm peak}.
\end{cases},
\]

and 

\[
S(\tau)\,=\,{\rm log}\,L_{\rm bol} (\tau) - {\rm log}\,L_{\rm peak}.
\]

\subsection{General Properties}

\begin{figure*}[t]
\includegraphics[width=0.8\textwidth]{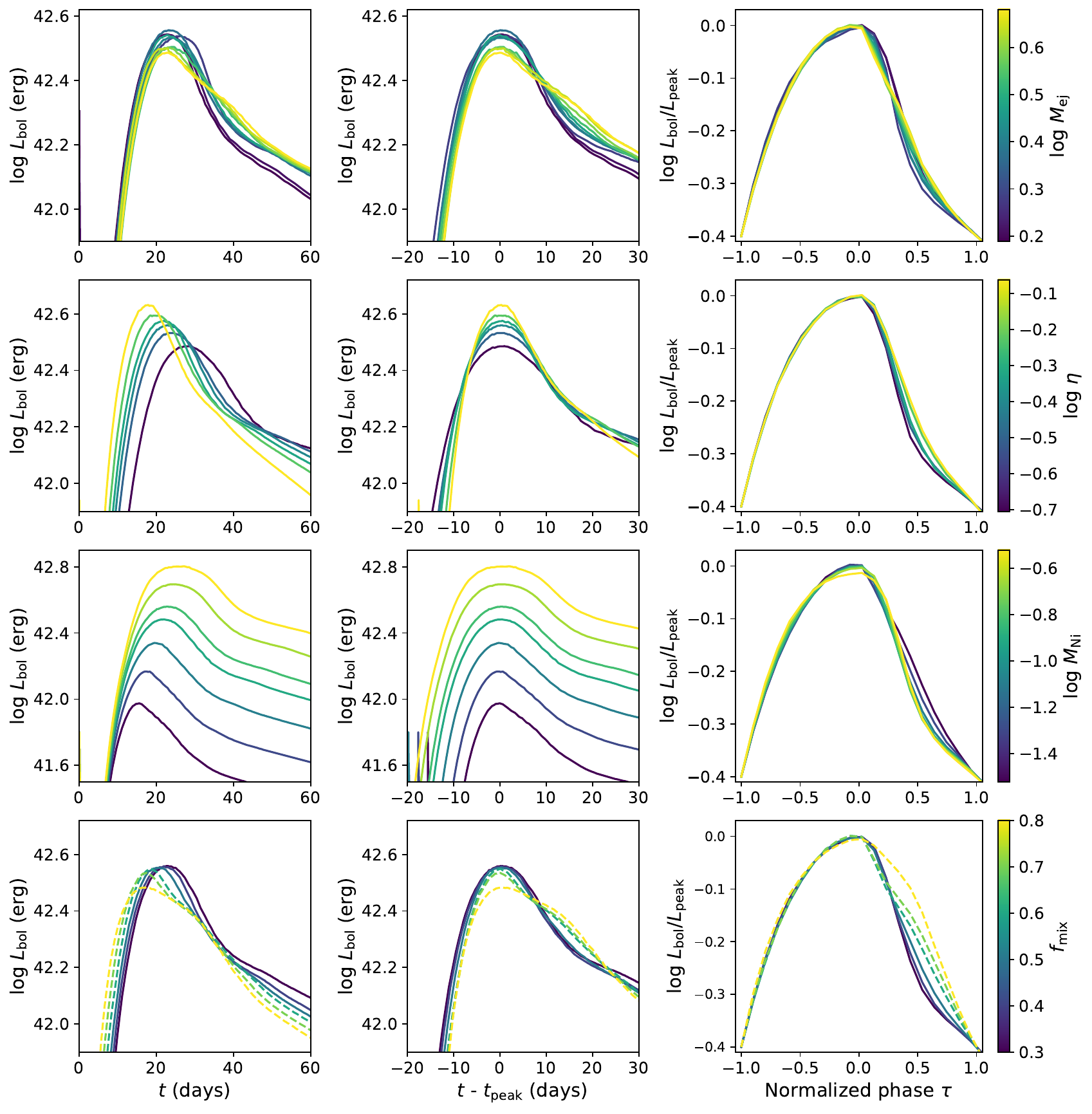}
\centering
\caption{The influences of the 4 parameters ($M_{\rm ej}$, $\eta$, $M_{\rm Ni}$ and $f_{\rm mix}$ from top to down) on the bolometric light curve since explosion (left panels), since peak time (middle panels) and light curve shape as function of normalized phase (right panels).}
\label{fig:sample_survey}
\end{figure*}

In Figure~\ref{fig:sample_survey}, we illustrate how the four parameters, i.e., $M_{\rm ej}$, $\eta$, $M_{\rm Ni}$ and $f_{\rm mix}$, affect the light curve properties. The default values are $M_{\rm ej}$\,=\,2.537\,$M_{\rm \odot}$ (corresponding to $M_{\rm ZAMS}$\,=\,15\,$M_{\rm \odot}$), $\eta$\,=\,0.394 (corresponding to $E_{\rm K}$\,=\,$10^{51}$\,erg for $M_{\rm ej}$\,=\,2.537\,$M_{\rm \odot}$), $M_{\rm Ni}$\,=\,0.15\,$M_{\rm \odot}$ and $f_{\rm mix}$\,=\,0.3. From top to bottom, each panel varies one parameter from its default value while keeping the other three fixed.

\begin{itemize}
    \item $M_{\rm ej}$. Increasing the ejecta mass increases $t_{\rm peak}$, $t_{\rm rise}$, and $t_{\rm decline}$, while simultaneously decreasing $L_{\rm peak}$. In addition, even at fixed $\eta$, models with relatively large $M_{\rm ej}$ exhibit a flatter post-peak evolution, sometimes producing a shoulder-like feature. Although this behavior may help distinguish models at the extreme ends of the parameter space, as shown later, a similar feature can also be reproduced by increasing the degree of mixing.

    \item $\eta$. Increasing $\eta$, or equivalently increasing $E_{\rm K}$ at fixed $M_{\rm ej}$, leads to a faster light-curve evolution, shortening all characteristic time scales while increasing $L_{\rm peak}$.

    \item $M_{\rm Ni}$. Increasing $M_{\rm Ni}$ enhances the luminosity at all phases as expected, while all timescales ($t_{\rm peak}$, $t_{\rm rise}$ and $t_{\rm decline}$) also increase. This behavior is not predicted by Arnett model, in which the shape of the light curve is only determined by $M_{\rm ej}$ and $v_{\rm ej}$, and varying $M_{\rm Ni}$ only changes $L_{\rm peak}$. The implication for this difference will be discussed in \S5.

    \item $f_{\rm mix}$. Increasing $f_{\rm mix}$ decreases $t_{\rm peak}$, as the $\gamma$-ray photons produced by radioactive decay require less time to diffuse out of the ejecta. For weakly mixed models, the light-curve properties around peak are only marginally affected, consistent with the findings of \citet{yadavalli26}. However, for strongly mixed models ($f_{\rm mix} > 0.5$ for the fiducial models; indicated by the dashed lines in the lower panels of Figure~\ref{fig:sample_survey}), $L_{\rm peak}$ decreases with increasing $f_{\rm mix}$. This is likely because the radioactive material is distributed closer to the surface, reducing the trapping efficiency of $\gamma$-ray photons. The light-curve morphology is also altered, with a shoulder-like feature emerging after peak, similar to the behavior seen in models with increasing $M_{\rm ej}$. A comparable effect was also reported by \citet{yadavalli26} for models in which $^{56}$Ni is efficiently mixed into the outer envelope (see their Figure~10).

\end{itemize}

In summary, the light-curve properties, including $L_{\rm peak}$ and all characteristic time scales, are affected by the four parameters in a complicated way. In contrast, the normalized light-curve shape, $S(\tau)$, is considerably less sensitive to parameter variations, as shown in the right panels of Figure~\ref{fig:sample_survey}. Once normalized by $t_{\rm rise}$, the rising phase ($\tau < 0$) is nearly identical across all models, while most of the diversity emerges during the declining phase ($\tau > 0$). This smooth and coherent behavior motivates the interpolation method introduced in Section~4.

\subsection{Scaling Relations}
Here we investigate the scaling relations between light curve properties and the physical parameters.

One of the most commonly employed method to estimate $M_{\rm Ni}$ from the light curves of hydrogen-poor CCSNe is the Arnett rule, which equals the luminosity of the light curve peak and the instantaneous radioactive decay energy, i.e.,

\begin{equation}
\begin{aligned}
L_{\rm peak}/M_{\rm Ni} &=
(\epsilon_{\rm Ni}-\epsilon_{\rm Co})
e^{-t_{\rm peak}/t_{\rm Ni}} \\
&\quad +
\epsilon_{\rm Co}
e^{-t_{\rm peak}/t_{\rm Co}}. 
\end{aligned}
\label{Eq:Arnett_Ni}
\end{equation}

\begin{figure}
\epsscale{1}
\plotone{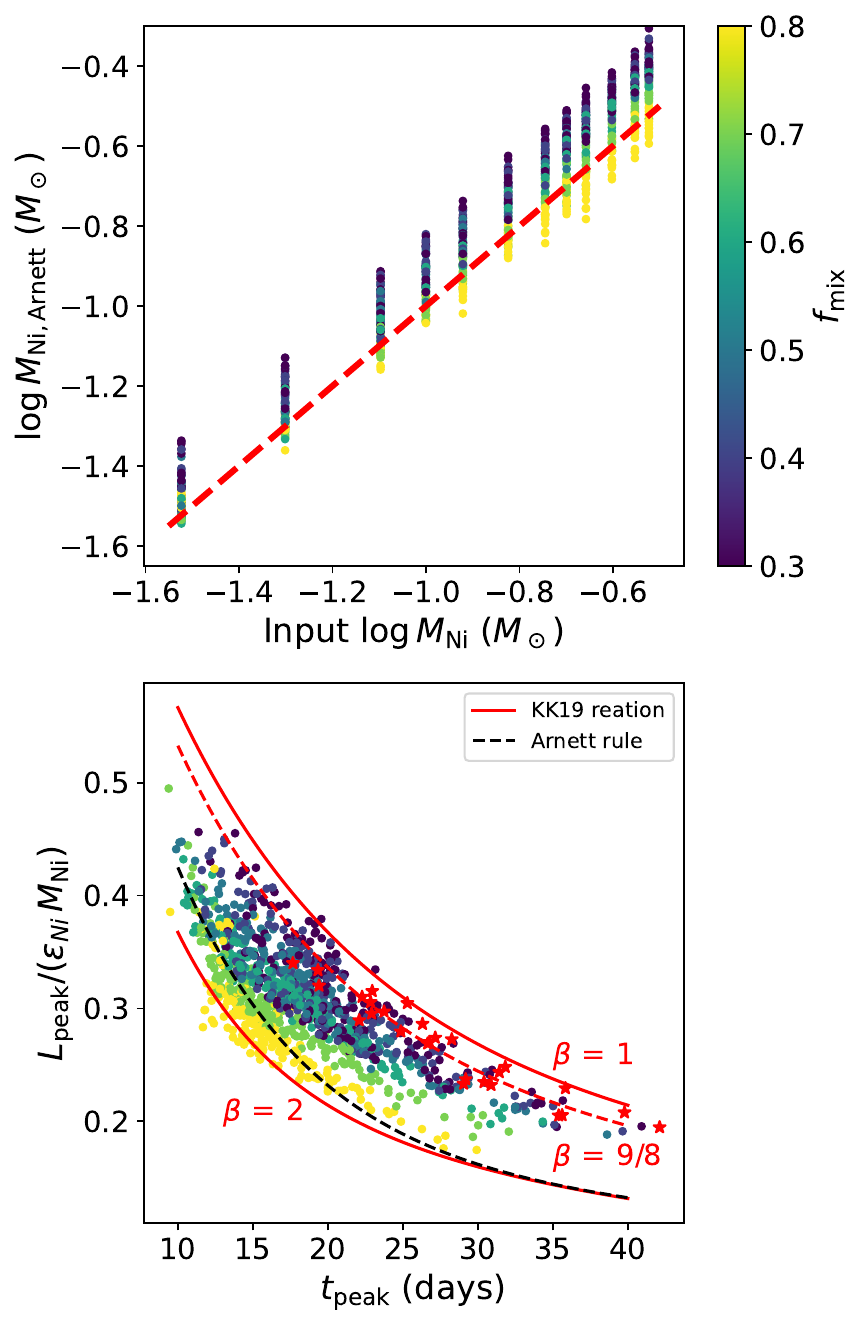}
\centering
\caption{Upper panel: comparison between the input $M_{\rm Ni}$ of the models with $M_{\rm Ni, Arnett}$ estimated based on the Arnett rule, color-coded by $f_{\rm mix}$; Lower panel: Comparison between $t_{\rm peak}$-$L_{\rm peak} $ and the relation of Eq.~\ref{Eq:KK19_beta} with different $\beta$ values (red lines) and the Arnett rule (black line). The red stars are the models from \citet{dessart16}.}
\label{fig:Scalings}
\end{figure}

Here $\epsilon_{\rm Ni}$ (=\,3.9\,$\times$\,10$^{10}$\,erg\,g$^{-1}$\,s$^{-1}$) and $\epsilon_{\rm Co}$ (=\,6.8\,$\times$\,10$^{9}$\,erg\,g$^{-1}$\,s$^{-1}$) are the heating rate per unit mass of $^{56}$Ni and $^{56}$Co, and $t_{\rm Ni}$ (=\,8.8\,days) and $t_{\rm Co}$ (=\,111.3\,days) are their decay time scales. With Equation~\ref{Eq:Arnett_Ni}, $M_{\rm Ni, Arnett}$ of the models are computed and compared with the input $M_{\rm Ni}$ of the numerical models, which are shown in the upper panel of Figure~\ref{fig:Scalings}. 

As is evident, $M_{\rm Ni, Arnett}$ overestimates the model $M_{\rm Ni}$ in nearly all cases, and the discrepancy becomes larger as $f_{\rm mix}$ decreases. This behavior arises because $L_{\rm peak}$ is largely insensitive to $f_{\rm mix}$ when the mixing is weak, and even decreases for strongly mixed models, while $t_{\rm peak}$ systematically decreases with increasing $f_{\rm mix}$. Together, these effects lead Equation~\ref{Eq:Arnett_Ni} to predict a larger $M_{\rm Ni}$ for more weakly mixed models. Consequently, the Arnett rule increasingly over predicts the $M_{\rm Ni}$. The discrepancy can reach $\sim$\,0.2\,dex (58\% in linear scale), comparable to \citet{dessart16}.

To overcome the limitation of Arnett rule, \cite{KK19} proposed a new relation between $t_{\rm peak}$ and $L_{\rm peak}$,

\begin{equation}
L_{\rm peak}\,=\,\frac{2}{(\beta t_{\rm peak})^2}\int_{0}^{\beta t_{\rm peak}} \tau L_{\rm heat}(\tau)\,d\tau, 
\label{Eq:KK19}
\end{equation}
where $L_{\rm heat}(t)$ is the heating function of the central source, and $\beta$ is a dimensionless parameter around unity that is dependent on many factors, including the recombination temperature, the concentration (or degree of mixing), etc, and can be calibrated with model light curves. For the radioactive decay chain, the above equation can be written as
\begin{equation}
\begin{aligned}
\frac{L_{\rm peak}}{\epsilon_{\rm Ni} M_{\rm Ni}} &= 2\,(\frac{t_{\rm Ni}}{\beta t_{\rm peak}})^2 \\
&\quad
[ (1 - \frac{\epsilon_{\rm Co}}{\epsilon_{\rm Ni}})\,(1 - (1 + \frac{\beta t_{\rm peak}}{t_{\rm Ni}})e^{-{\beta t_{\rm peak}/t_{\rm Ni}}}) \\
&\quad +
\frac{\epsilon_{\rm Co}t_{\rm Co}^2}{\epsilon_{\rm Ni}t_{\rm Ni}^2}
\,(1 - (1 + \frac{\beta t_{\rm peak}}{t_{\rm Co}})e^{-{\beta t_{\rm peak}/t_{\rm Co}}})]. 
\end{aligned}
\label{Eq:KK19_beta}
\end{equation}

In the lower panel of Figure 2, we estimate $\beta$ by comparing $t_{\rm peak}$ and $L_{\rm peak}$ of the models, and find almost all of them fall within the range of $\beta$\,=\,1 to $\beta$\,=\,2. Specially, $\beta$ increases with $f_{\rm mix}$, consistent with the results of \cite{KK19}, where they estimated
\[
\beta\,\approx\,\frac{4}{3}(1 + x_{\rm s}^{4}).
\]
Here $x_{\rm s}$ is the spatial concentration, which is defined on radial coordinate and different from $f_{\rm mix}$ defined on mass coordinate, but the quantitative behaviors are the same: strongly mixed models have larger $\beta$.

Linear regression on the logarithmic scale yields
\[
\frac{M_{\rm Ni}}{M_{\rm \odot}}\,=\,0.04\,(\frac{L_{\rm peak}}{10^{42}\,{\rm erg\,s^{-1}}})^{1.01}(\frac{t_{\rm peak}}{20\,{\rm days}})^{0.52}.
\]
The standard deviation of the residuals is 0.04\,dex (10\% in linear scale). If we restrict the fit to models with $f_{\rm mix}\,\leq\,$0.5, i.e., at most half of the ejecta is mixed, the relation becomes
\[
\frac{M_{\rm Ni}}{M_{\rm \odot}}\,=\,0.04\,(\frac{L_{\rm peak}}{10^{42}\,{\rm erg\,s^{-1}}})^{0.98}(\frac{t_{\rm peak}}{20\,{\rm days}})^{0.65},
\]
and the standard deviation of the residuals reduced to 0.03\,dex (7\% in linear scale).

However, in most cases, $t_{\rm peak}$ is weakly constrained by observation, as its determination requires high-cadence data between the non-detections and the first detection. More importantly, if $^{56}$Ni is deeply buried within the ejecta and the progenitor lacks an extended envelope capable of producing shock-cooling emission (like SNe IIb), the SN may experience a "dark period" lasting up to $\sim$\,10--20 days and more than 2 dex (5 mag) fainter than the peak before the light curve begins to rise. Consequently, even a non-detection does not necessarily imply that the explosion has not yet occurred.

Similarly to \citet{yadavalli26}, we employ the properties around the peak, i.e., $t_{\rm rise}$ and $t_{\rm decline}$, instead of $t_{\rm peak}$ in the fitting, which yield

\[
\frac{M_{\rm Ni}}{M_{\rm \odot}}\,=\,0.04\,(\frac{L_{\rm peak}}{10^{42}\,{\rm erg\,s^{-1}}})^{1.03}(\frac{t_{\rm rise}}{15\,{\rm days}})^{0.25}(\frac{t_{\rm decline}}{20\,{\rm days}})^{0.33},
\]
with the standard deviation of the residuals reduced to 0.01 dex (2\% in linear scale). The result is broadly consistent with \citet{yadavalli26}. Unlike relations based on $t_{\rm peak}$, it relies solely on observables measured around maximum light and is thus less affected by the uncertainties associated with the loosely constrained explosion epoch.

\section{Interpolation-Based Inference Framework}
In this section, we develop an interpolation-based inference framework that could turn the discrete model grid into a continuous estimator, which would be applied to realistic observation in forthcoming works.

\subsection{Interpolation}
Before the model grid can be applied to observed SNe, it is necessary to predict light curves at arbitrary locations within the parameter space. Because the grid is sampled only at a finite number of parameter combinations, we employ an interpolation scheme to estimate light curves between grid points.

For an arbitrary target parameter set \{log\,$M_{\rm ej,0}$, log\,$\eta_{0}$, log\,$M_{\rm Ni,0}$, $f_{\rm mix,0}$\}, we first identify the indices \{im, i, j, k\} that bracket the target model in parameter space, such that
\[
{\rm log}\,M_{\rm ej}[{\rm im}] \leq {\rm log}\,M_{\rm ej,0} < {\rm log}\,M_{\rm ej}[{\rm im+1}].
\]
We then define the normalized coordinate (or relative distance) along each parameter dimension. For example, in the log\,$M_{\rm ej}$ direction,
\[
u({\rm log\,M_{\rm ej}}) = \frac{{\rm log}\,M_{\rm ej,0} - {\rm log}\,M_{\rm ej}[{\rm im}]}{{\rm log}\,M_{\rm ej}[{\rm im + 1}] - {\rm log}\,M_{\rm ej}[{\rm im}]}, 
\]
with analogous definitions for log\,$\eta$, log\,$M_{\rm Ni}$, and $f_{\rm mix}$. The interpolated light curve is obtained through four-dimensional linear interpolations for the light curve shape $S(\tau)$, log\,$L_{\rm peak}$, log\,$t_{\rm peak}$, log\,$t_{\rm rise}$ and log\,$t_{\rm decline}$, in which the (2$^4$=16) models surrounding the target point are combined with weights determined by their relative distances along each parameter dimension. The physical light curve is then reconstructed using the relations described in \S~3.

To assess both the smoothness of the model grid and the accuracy of the interpolation scheme, we perform a stringent cross-validation test. For each model in the grid, we temporarily remove all models that share any parameter value with the target model. For example, for a target model with \{${\rm log}\,M_{\rm ej},{\rm log}\,\eta,{\rm log}\,M_{\rm Ni},f_{\rm mix}$\}\,=\,\{0.4,-0.4,-1.0,,0.7\}, we exclude all models with log\,$M_{\rm ej}$\,=\,0.4, all models with log\,$\eta$\,=\,0.4, all models with log\,$M_{\rm Ni}$\,=\,-1.0, and all models with $f_{\rm mix}$\,=\,0.7. The target model is then reconstructed using interpolation from the remaining grid points.

This procedure ensures that no model used in the interpolation shares a parameter value with the target model, thereby eliminating the nearest neighbors in all four dimensions. The reconstructed light curve is subsequently compared with the original model, and the interpolation accuracy is quantified through the residuals between the two. Repeating this procedure for all interior grid models provides a stringent test of the robustness of the interpolation scheme.

\begin{figure}
\epsscale{1}
\plotone{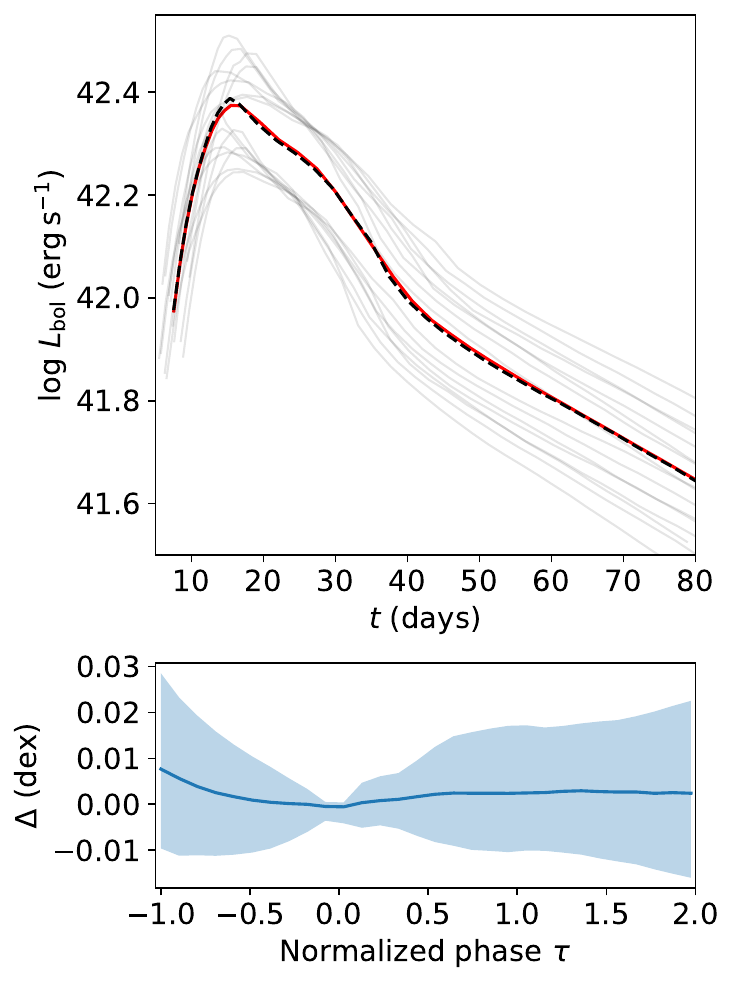}
\centering
\caption{Upper panel: Example of interpolation. The target models has \{$M_{\rm ej},\eta,M_{\rm Ni},f_{\rm mix}$\}\,=\,\{2.54,0.394,0.10,0.7\} (\{${\rm log}\,M_{\rm ej},{\rm log}\,\eta,{\rm log}\,M_{\rm Ni},f_{\rm mix}$\}\,=\,\{0.40,-0.42,-1.0,,0.7\}). The red solid line is the interpolated model and the black dashed line is the true model. The transparent lines are models bracketing the target model: $M_{\rm ej}$\,=\,\{2.34, 2.96\}, $\eta$\,=\,\{0.342, 0.513\} (for $M_{\rm ej}$\,=\,2.34), $\eta$\,=\,\{0.338, 0.508\} (for $M_{\rm ej}$\,=\,2.96), $M_{\rm Ni}$\,=\,\{0.08, 0.12\} and $f_{\rm mix}$\,=\,\{0.6, 0.8\}. All masses are in the unit of $M_{\rm \odot}$ and $\eta$ is in the unit of 10$^{51}$\,erg/$M_{\rm \odot}$. Lower panel: the residuals $\Delta$ as the function of $\tau$. The solid line is the median and the transparent region represents 68\% CI for the slice-removed interpolation tests.}
\label{fig:Interpolation}
\end{figure}

An example of this validation test is shown in the upper panel of Figure~\ref{fig:Interpolation}. The lower panel summarizes the interpolation accuracy by showing the median residual together with the 68\% confidence interval (CI), defined by the 16$^{\rm th}$ and 84$^{\rm th}$ percentiles. The residual is computed in logarithmic luminosity space,

\[
\Delta = \log L_{\rm model}(t) - \log L_{\rm interp}(t),
\]
where $L_{\rm model}$ and $L_{\rm interp}$ denote the original and interpolated light curves, respectively. Although computed in phase space, to allow for a uniform comparison throughout the model grid, the residuals are presented as a function of the normalized phase $\tau$ of the target model.

The interpolation is essentially unbiased, with the median residual remaining below 0.01\,dex throughout all phases, corresponding to a systematic offset of less than $\sim$\,1\%. The scatter exhibits a modest phase dependence. Near maximum light ($\tau\approx0$), the 68\% CI is approximately $\pm$\,0.003 dex. The scatter increases away from the peak, reaching $\sim$\,0.02\,dex at $\tau\,=$\,2 and $\sim\,0.02$ dex (-0.010 dex to +0.028 dex) at $\tau\,=$\,-1.

These scatter levels should be regarded as a conservative upper limit because the validation procedure removes all models that share any parameter value with the target model. In practical applications, the interpolation is performed on the full grid with a denser local sampling. We therefore adopt $\sigma_{\rm interp}$\,=\,0.02\,dex as a conservative estimate of the uncertainty of interpolation.

\subsection{Application to Observations}

After the interpolation framework is established, we employ it to infer the physical properties of two well-observed SNe Ib: SNe 2007Y and SNe 2009jf. These two objects sit at two extremes of the helium core masses of the SNe Ib progenitors, as reflected by the relative intensities of their [O I] lines seen in nebular phase spectroscopy \citep{fang19,fang22}, therefore can be employed to test whether the model grid can produce the observed light curves.

The photometry, as well as the distances and extinctions, are derived from the literature (SN 2007Y: \citealt{stritzinger09}; SN 2009jf: \citealt{valenti11}). The bolometric light curves are constructed using \texttt{Superbol} \citep{nicholl18}. 

The observed light curves are then fitted using the interpolation-based Markov chain Monte Carlo (MCMC) framework implemented with \texttt{emcee} \citep{emcee}. Prior to the fitting, both the observations and the model light curves are shifted to their respective peak epochs. The likelihood function is defined as:
\[
{\rm ln}\,\mathcal{L}\,=-\frac{1}{2}\,\sum\frac{[{\rm log}L_{\rm obs, obs} - {\rm log}L_{\rm model}(\theta)]^2}{\sigma_{\rm obs}^2 + \sigma_{\rm interp}^2} + C,
\]
where $\theta$ is an arbitrary combination of physical parameters within their ranges and $L_{\rm model}(\theta)$ is the luminosity of the interpolated light curve. Here, $\sigma_{\rm interp}$(0.02\,dex) represents the conservative interpolation uncertainty estimated in \S4.1.

Uniform priors are adopted for all four physical parameters. In addition, we introduce a phase-shift parameter, $\Delta t_{\rm shift}$, to account for uncertainties in the peak alignment. A uniform prior between -5 and +5 days is assumed for this parameter. The MCMC chains are initialized with 50 walkers (10 per dimension), randomly distributed according to the adopted priors, and evolved until convergence is achieved.

\begin{figure*}[htbp]
    \centering
    \begin{minipage}[t]{0.45\textwidth}
        \begin{overpic}[width=\textwidth]{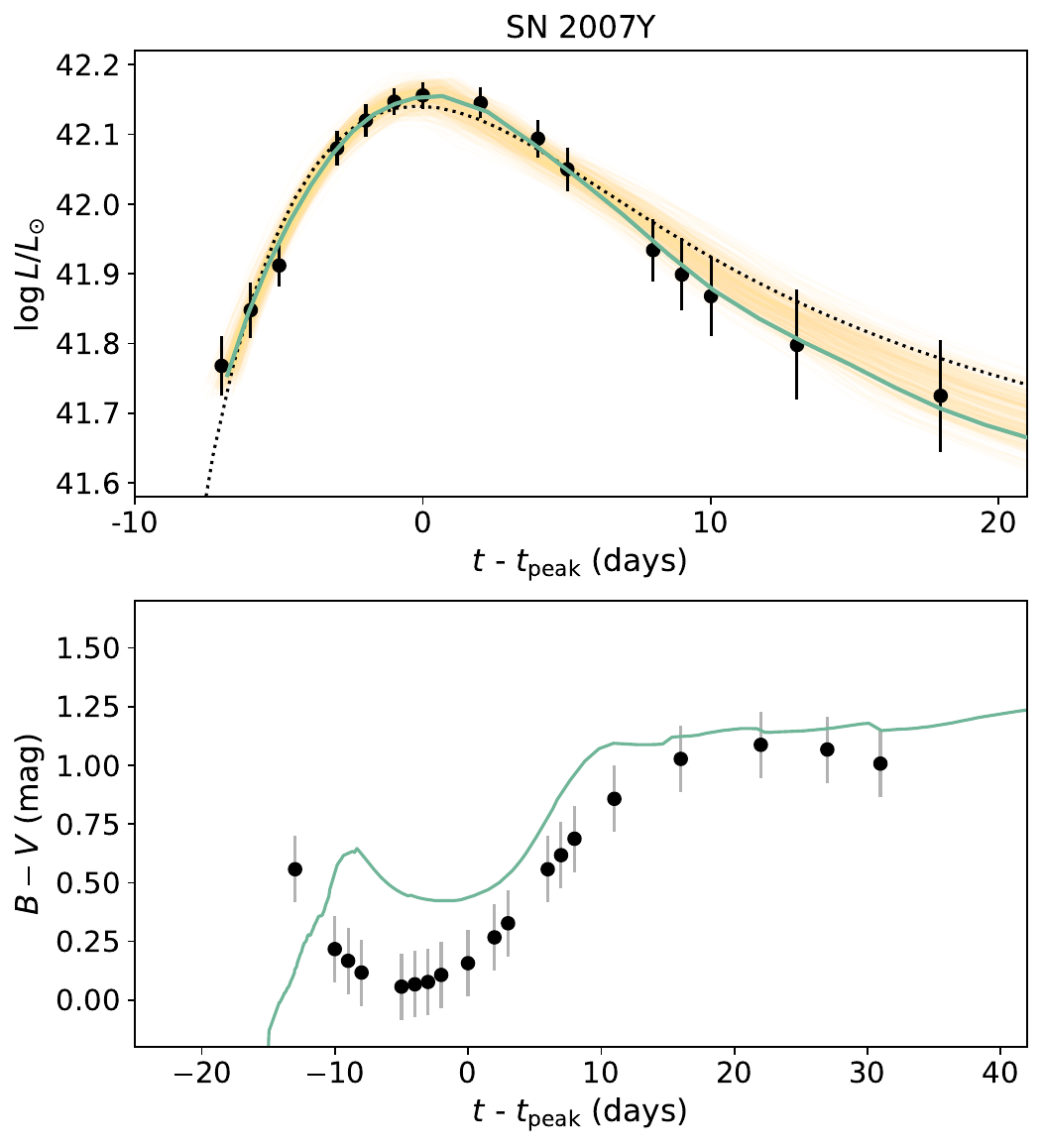}
        \end{overpic}
    \end{minipage}%
    \hfill
    \begin{minipage}[t]{0.45\textwidth}
        \begin{overpic}[width=\textwidth]{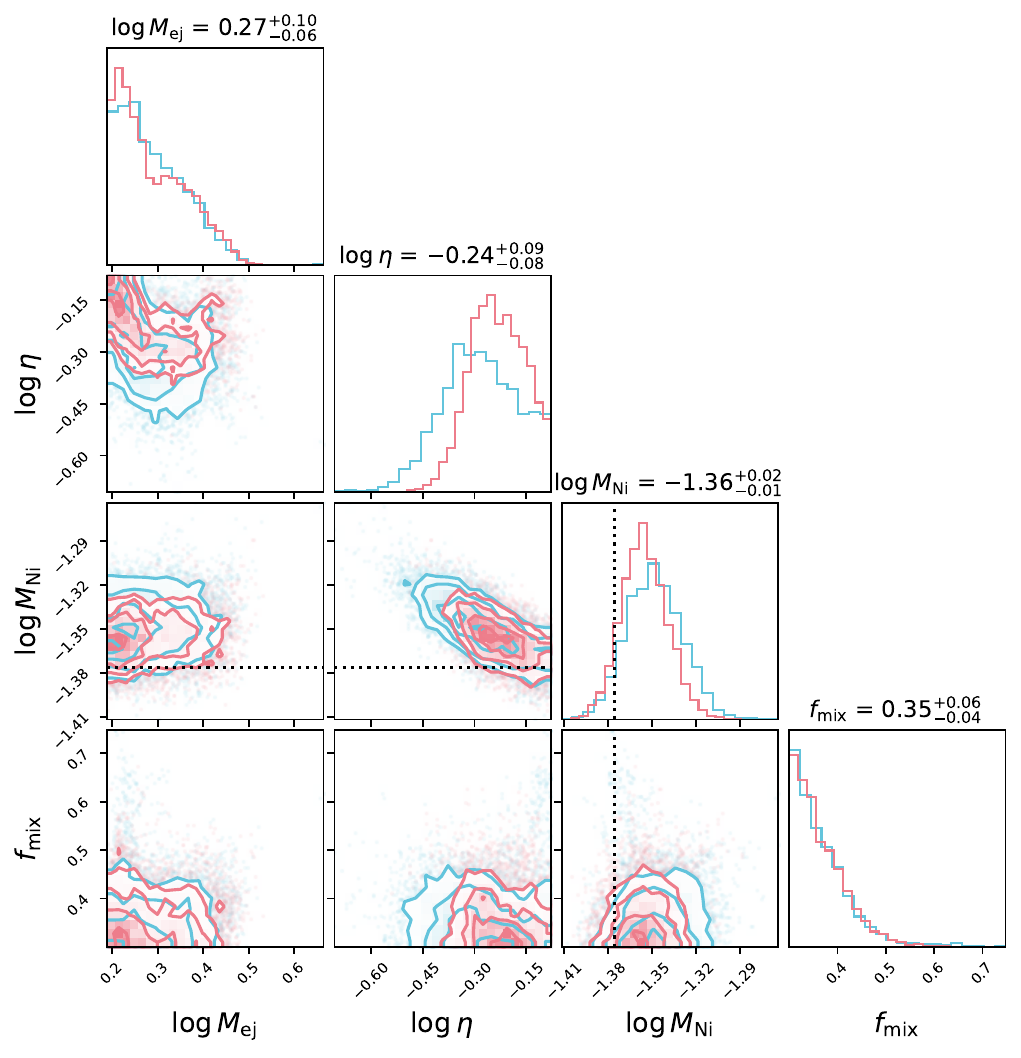}
        \end{overpic}
    \end{minipage}

    \caption{Parameter inference for SN 2007Y. Upper left panel: the bolometric light curve of observation (black scatter points), the optimized model (green solod line) and 300 models randomly drawn from the posterior (yellow transparent lines). Lower left panel: the $B-V$ color of SN 2007Y (black scatter points) and the optimized model (green solid line). Right panels: the posteriors with (pink) and without (lightblue) the characteristic velocity as prior. In both panels, the black dotted lines represent the optimized values of the analytical light curve models.}
    \label{fig:SN07Y}
\end{figure*}

\begin{figure*}[htbp]
    \centering
    \begin{minipage}[t]{0.45\textwidth}
        \begin{overpic}[width=\textwidth]{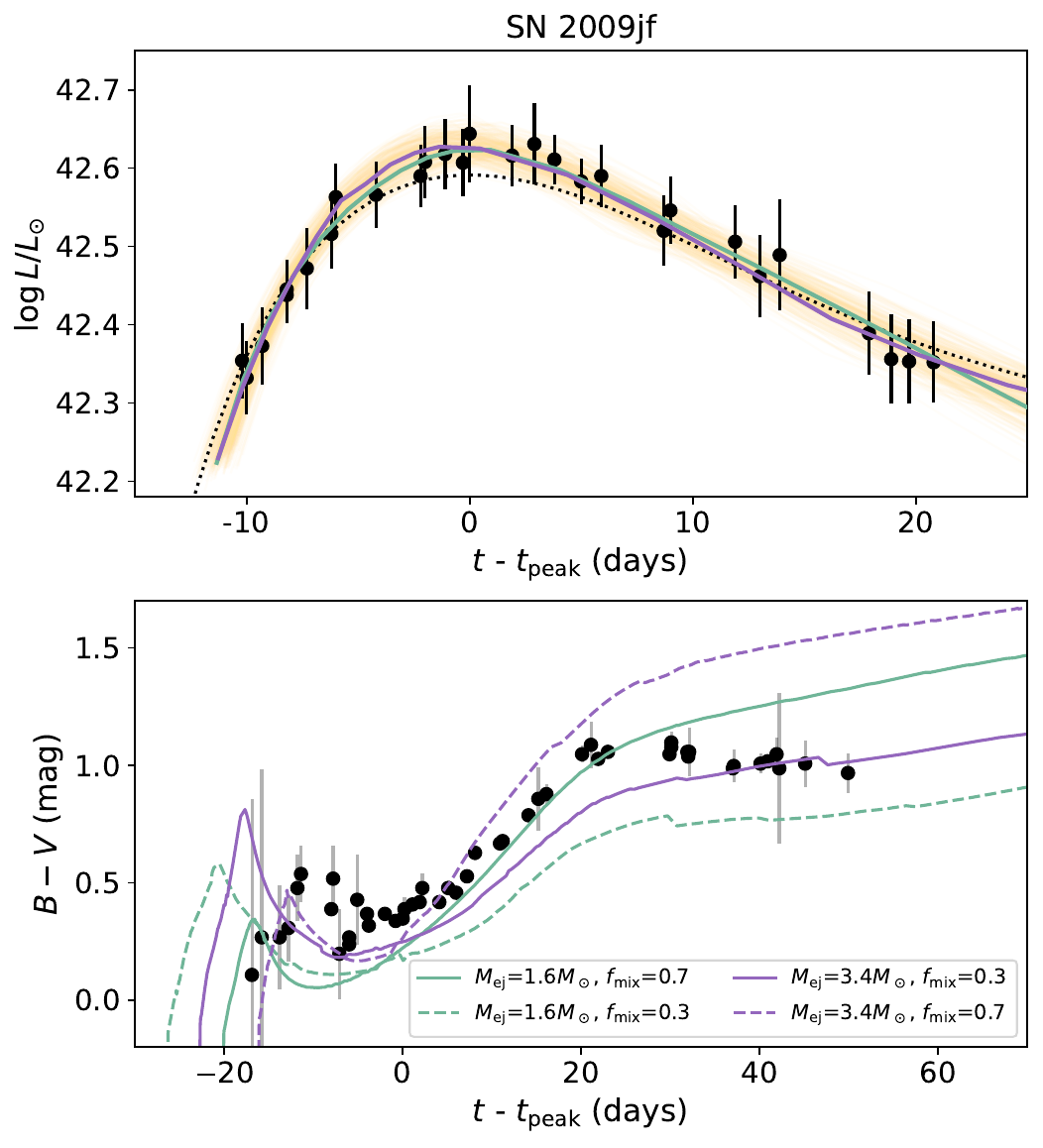}
        \end{overpic}
    \end{minipage}%
    \hfill
    \begin{minipage}[t]{0.45\textwidth}
        \begin{overpic}[width=\textwidth]{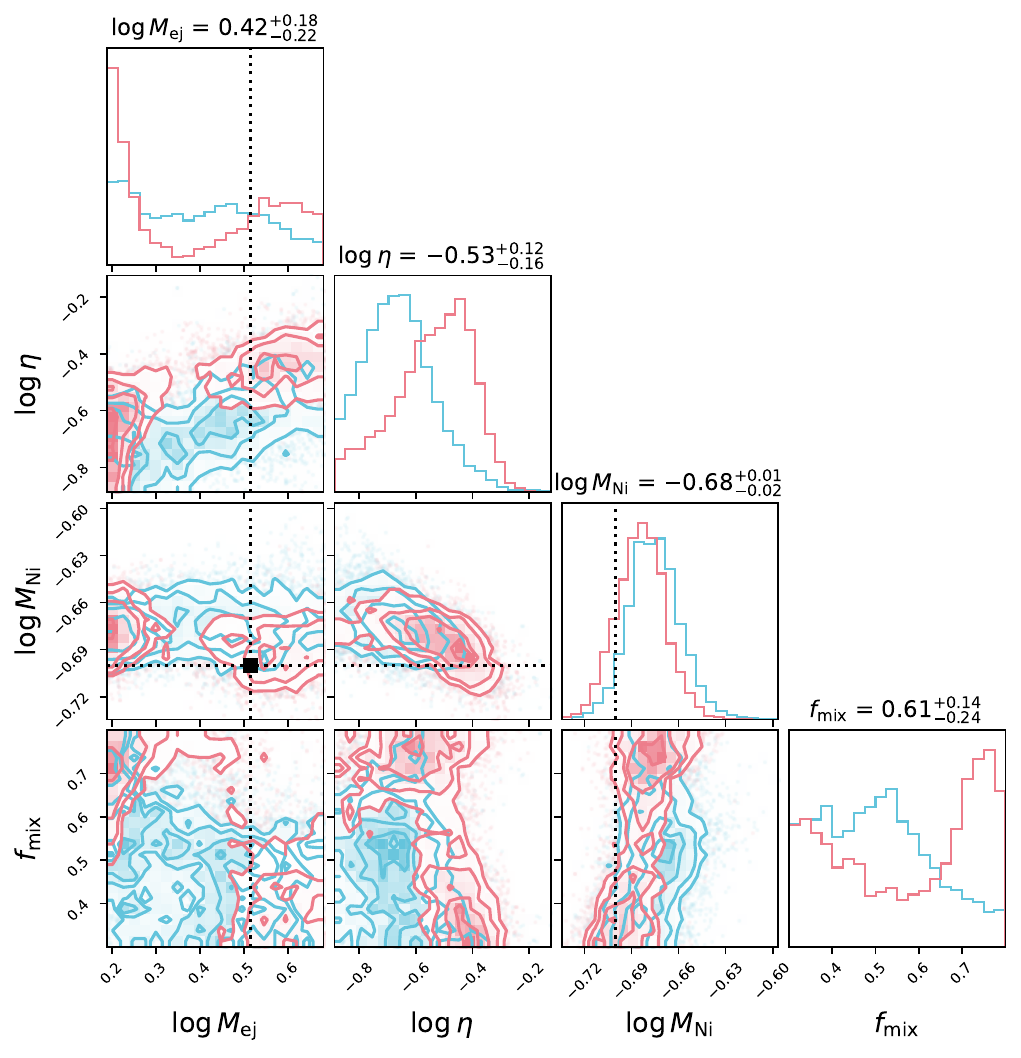}
        \end{overpic}
    \end{minipage}

    \caption{Same as Figure~\ref{fig:SN07Y} but for SN 2009jf. In the left panels, the green and purple solid lines represent the optimized model at the lower-mass and the higher-mass branches, respectively.}
    \label{fig:SN09jf}
\end{figure*}

\begin{itemize}
    \item SN 2007Y.~~~The posteriors distributions and the optimized model light curves are shown in Figure~\ref{fig:SN07Y}. The inferred parameters are: log\,$M_{\rm ej}$ = $0.27^{+0.09}_{-0.05}$ ($M_{\rm ej}$ = 1.86$^{+0.43}_{-0.21}$\,$M_{\rm \odot}$), log\,$\eta$ = $-0.29^{+0.13}_{-0.11}$ ($E_{\rm K}$ = $1.00^{+0.28}_{-0.21}$\,foe), log\,$M_{\rm Ni}$ = $-1.35^{+0.02}_{-0.02}$ ($M_{\rm Ni}$ = 0.045$^{+0.002}_{-0.002}$\,$M_\odot$), and $f_{\rm mix}$ = $0.35^{+0.07}_{-0.04}$. Using the mapping presented in Table~\ref{tab:grid_params}, these parameters correspond to a progenitor with $M_{\rm ZAMS}\sim$13\,$M_{\rm \odot}$, consistent with the relatively weak [O~I] emission observed in the nebular spectra. However, this mapping is subject to the systematic uncertainties discussed in \S2.

    Interestingly, the model predicts a Fe~II $\lambda5169$ velocity\footnote{Defined as the velocity of the mass shell with Sobolev opacity $\tau$=1 for Fe~II $\lambda5169$.} of $8380^{+750}_{-660}$\,km\,s$^{-1}$ at peak light, in good agreement with the observed absorption minimum of $9000$\,km\,s$^{-1}$ (\citealt{lyman16}). Furthermore, including the observed Fe~II velocity as an additional prior in the MCMC analysis does not significantly alter the posterior distributions (right panels of Figure~\ref{fig:SN07Y}), suggesting that the light-curve constraints already provide a consistent estimate of the ejecta properties for this object.

    For comparison, we also fit the data with the Arnett-type analytical light-curve model (see \S5 for details), obtaining log\,$M_{\rm ej}=-0.18^{+0.08}_{-0.08}$ ($M_{\rm ej}=0.65^{+0.13}_{-0.11}$\,$M_{\rm \odot}$) and log\,$M_{\rm Ni}=-1.38^{+0.017}_{-0.015}$ ($M_{\rm Ni}=0.042^{+0.002}_{-0.001}$\,$M_{\rm \odot}$). Compared with the parameters inferred from the numerical models, the analytical model yields a lower $M_{\rm ej}$, although this discrepancy may partly arise from the limited parameter range covered by our numerical model grid.

    In the lower-left panel of Figure~\ref{fig:SN07Y}, we compare the $B-V$ curve of SN 2007Y with that of the optimized model, represented by the broadband light curves of the nearest grid point in parameter space. The observed color evolution exhibits two distinct phases: an initial blueward evolution driven by the delayed heating of deeply buried $^{56}$Ni, followed by a redward evolution toward the nebular phase. These behaviors have been suggested by \citet{yoon19} as characteristic of SESNe in which $^{56}$Ni is only weakly mixed. The initial reddening phase is not observed in SN 2007Y because of the lack of sufficiently early-time data. The model exhibits similar overall behavior, although the color trough is shallower than observed, possibly due to the relatively simplified mixing prescription adopted in this work. Nevertheless, the model successfully reproduces the nearly saturated color at later phases.

    \item SN 2009jf.~~~~The posteriors distributions and the optimized model light curves are shown in Figure~\ref{fig:SN09jf}. Unlike SN 2007Y, where the parameters are quite well-constrained, the posteriors of SN 2009jf show large scatters, and $M_{\rm ej}$ exibits an almost flat distribution. Importantly, when the observed Fe~II velocity is included as an prior (9500\,km\,s$^{-1}$;\citealt{valenti11,lyman16}), the $M_{\rm ej}$ and $f_{\rm mix}$ posteriors become bimodal, with a lower-mass branch ($M_{\rm ej}\sim1.6\,M_\odot$) characterized by high $f_{\rm mix}\sim0.7$ and a higher-mass branch ($M_{\rm ej}\sim3.4\,M_\odot$) characterized by low $f_{\rm mix}\sim0.3$. This bimodality reflects the degeneracy between $M_{\rm ej}$ and $f_{\rm mix}$ discussed in \S2. As shown in the upper-left panel of Figure~\ref{fig:SN09jf}, the optimized models corresponding to the two branches produce nearly identical bolometric light curves.

    Similarly, the fit with the Arnett-type analytical light-curve model (see \S5 for details) gives log\,$M_{\rm ej}=0.52^{+0.26}_{-0.32}$ ($M_{\rm ej}=3.27^{+2.74}_{-1.71}$\,$M_{\rm \odot}$) and log\,$M_{\rm Ni}=-0.704^{+0.135}_{-0.101}$ ($M_{\rm Ni}=0.198^{+0.072}_{-0.041}$\,$M_{\rm \odot}$). The inferred $M_{\rm ej}$ is more consistent with the higher-mass branch (see \S5 for a systematic comparison between numerical and analytical models).

    In the lower left panel of Figure~\ref{fig:SN09jf}, we compare the $B-V$ curves of SN 2009jf and those of the optimized models of the low-mass (green solid line) and high-mass (purple solid line) branches. For each branch, the color evolution is represented by the broadband light curves of the grid model nearest to the inferred parameters in parameter space. The high-mass model provides a better match to the observed pre-peak and late-time saturated colors, whereas the low-mass model shows better agreement during the post-peak cooling phase. However, given the uncertainties in the host-galaxy extinction and those associated with the LTE assumption adopted in \texttt{STELLA}, it is difficult to determine which model provides a more satisfactory overall description of the observed color evolution.

    This comparison also illustrates that the color evolution alone provides limited constraints on $f_{\rm mix}$. To demonstrate this point, we show in the lower-left panel of Figure~\ref{fig:SN09jf} the color curves of two additional models: a low-mass model with $f_{\rm mix}=0.3$ (green dashed line) and a high-mass model with $f_{\rm mix}=0.7$ (purple dashed line). These models are identical to the corresponding optimized models shown by the solid lines, except for their values of $f_{\rm mix}$. If all other parameters are fixed, the late-time saturated color does provide a clear constraint on $f_{\rm mix}$: the weakly mixed models are bluer than the strongly mixed models by approximately 0.5\,mag in $B-V$. However, the degeneracy in the bolometric light curve induces an anti-correlation between $M_{\rm ej}$ and $f_{\rm mix}$. In our model grid, lower-$M_{\rm ej}$ models are generally bluer than higher-$M_{\rm ej}$ models (see also Figure~9 of \citealt{woosley21}, although the physical parameters in their models are not independent). This difference in color partially compensates for the reddening associated with stronger $^{56}$Ni mixing. Consequently, the color curves of the optimized models corresponding to the two branches are much more similar than those of models with fixed $M_{\rm ej}$, resulting in limited additional constraining power from the observed color evolution.

\end{itemize}

These examples demonstrate that the model grid, combined with the interpolation scheme, can successfully reproduce the observed bolometric light curves. In both cases, $^{56}$Ni mass is robustly constrained regardless of whether the characteristic velocity is available, whereas the inferred ejecta mass $M_{\rm ej}$, specific kinetic energy $\eta$, and mixing parameter $f_{\rm mix}$ remain poorly constrained (see also \citealt{sarin26}). In particular, the degeneracy between $M_{\rm ej}$ and $f_{\rm mix}$ persists even when additional information from the color evolution is included. In a forthcoming paper, we will apply this framework to a large sample of SESNe and develop additional methods to break these parameter degeneracies and obtain more robust constraints on their physical properties.

\section{Comparison with Arnett-model Inference}

Arnett-type models, initially proposed by \citet{arnett82}, provide analytical descriptions of SNe light curves powered predominantly by a central heating source and have been widely applied to observations of SESNe to infer progenitor and explosion properties. Despite their extensive use, the reliability of the inferred parameters has not been systematically assessed. In particular, it remains unclear to what extent the quantities derived from the fitting of the Arnett-model, such as the ejecta mass $M_{\rm ej}$, accurately represent the underlying physical parameters of the explosion. The magnitude of the associated systematic biases and uncertainties, and their impact on our interpretation of SESN progenitors, also remain poorly quantified. The primary goal of this section is to investigate these issues using the radiation-hydrodynamic model grid developed in this work. By comparing the parameters inferred from the Arnett model fitting with the known properties of the underlying models, we quantify both the accuracy and the limitations of the Arnett framework for recovering the physical characteristics of SESNe.

Arnett-type models rely on several simplified assumptions:
\begin{itemize}
\item The progenitor radius at the onset of the explosion is negligible;
\item The ejecta is already in the homologous expansion phase and radiation pressure dominates;
\item The optical opacity $\kappa$ and characteristic ejecta velocity $v_{\rm ej}$ remain constant throughout the evolution;
\item The temperature, velocity, and $^{56}$Ni distributions are related in a simplified way. As a consequence, the energy-density profile is self-similar and time-independent \citep{KK19}.
\end{itemize}
The emergent luminosity can then be solved from the diffusion equation. For SESNe, the heating source is generally assumed to arise entirely from the radioactive decay chain of $^{56}$Ni (Equation~\ref{Eq:Arnett_Ni}).

Once the functional form of $L_{\rm heat}$ is specified, the overall light-curve shape, including its rise and decline timescales, is only controlled by $t_{\rm d}$, which effectively measures the ratio $M_{\rm ej}/v_{\rm ej}$, assuming a fixed $\kappa$. With an independent constraint on the photospheric velocity, usually estimated from the absorption minimum of Fe~II $\lambda5169$ near maximum light, the degeneracy between $M_{\rm ej}$ and $v_{\rm ej}$ can in principle be removed.

To assess the extent to which these inferred quantities reflect the input physical parameters represented in our model grid, we compare the inferred parameters, based on the Arnett-type models, with the those of the synthetic light curves. The mock observations are generated from the models by assuming a typical cadence of 1 to 3 days between $-1<\tau <+1$ for the normalized phase, and adopting $\sigma_{\rm obs}=0.02$\,dex (corresponding to the uncertainty of bolometric correction; see \citealt{lyman14}) to perturb log\,$L_{\rm model}$. We employ the \texttt{arnett\_bolometric} model (\citealt{arnett82}) implemented in \texttt{Redback} (\citealt{sarin24}). The priors are listed in Table~\ref{tab:priors}. For simplicity, the optical opacity is fixed at $\kappa=0.06~{\rm cm^2\,g^{-1}}$ \citep{maeda03,valenti08}\footnote{The value of $\kappa$ is very uncertain (see, e.g., the discussions in \citealt{wheeler15}), and it can affect the outcome through degeneracy in $\kappa M_{\rm ej}$. However, to compare with the results of \citet{lyman16}, we strictly follow their methodology and apply a fixed $\kappa=0.06~{\rm cm^2\,g^{-1}}$.}. Note that the model uses the mass fraction of $^{56}$Ni, $f_{\rm Ni}$($\coloneqq M_{\rm Ni}$/$M_{\rm ej}$), instead of $M_{\rm Ni}$ as the major parameter.

\begin{figure*}
\epsscale{1}
\plotone{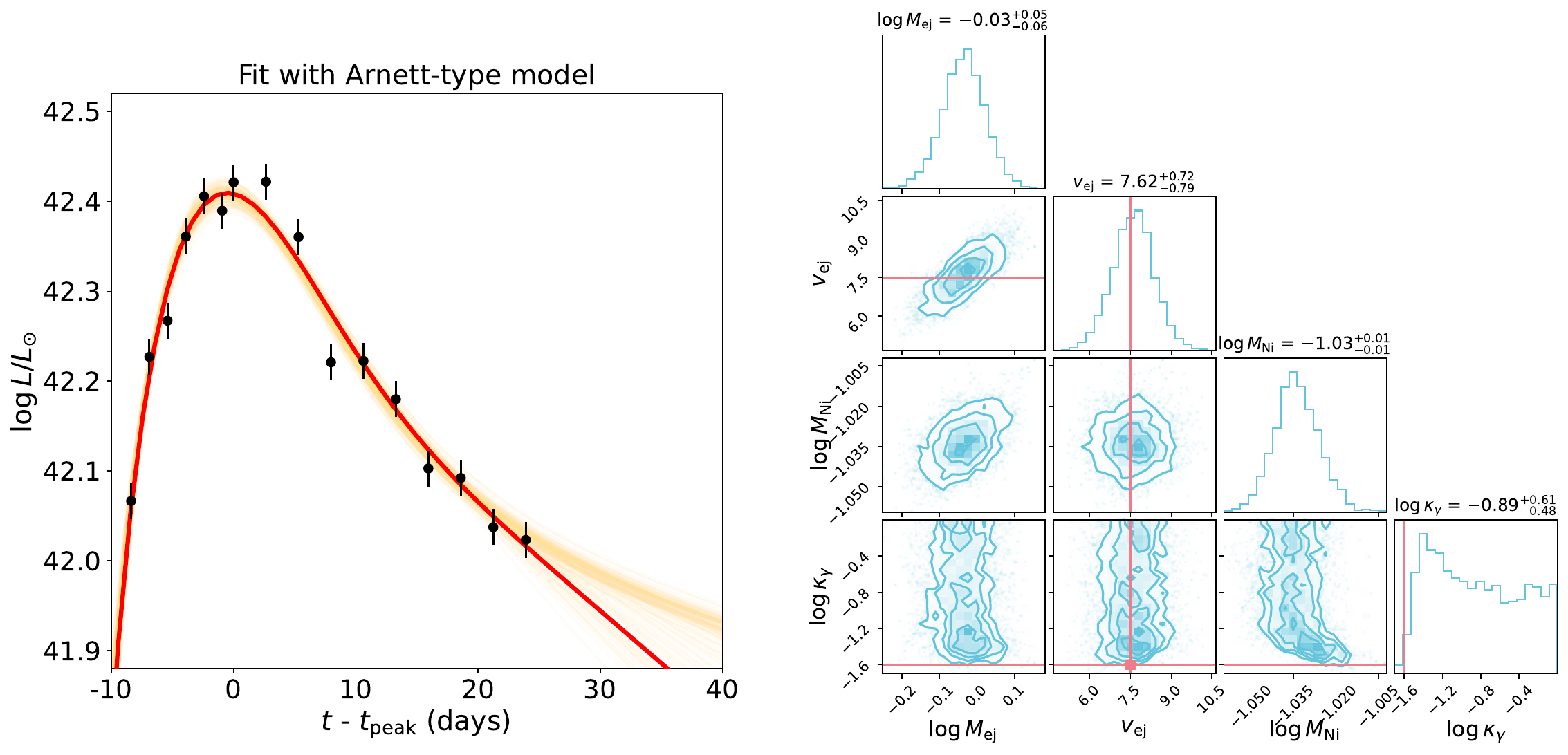}
\centering
\caption{Parameter inference using Arnett-type models. Left panel: The target models has \{$M_{\rm ej},\eta,M_{\rm Ni},f_{\rm mix}$\} = \{2.54,0.394,0.10,0.3\} (\{${\rm log}\,M_{\rm ej},{\rm log}\,\eta,{\rm log}\,M_{\rm Ni},f_{\rm mix}$\} = \{0.40,-0.42,-1.0,0.3\}). The scatter points are the mock observation generated from the radiation-hydrodynamic model with $\sigma_{\rm obs}\,=\,0.02\,$dex. The red solid line is the optimized Arnett-type model and the transparent lines are 300 models randomly drawn from the posterior distributions. Right panels: the posterior distributions and covariances of the parameters inferred from the Arnett-type model. The input values of log\,$M_{\rm ej}$ and log\,$M_{\rm Ni}$ (=$f_{\rm Ni}\,M_{\rm ej}$) fall outside the 3-$\sigma$ ranges of their corresponding posteriors.}
\label{fig:Inference_Arnett_example}
\end{figure*}

\begin{figure}
\epsscale{0.8}
\plotone{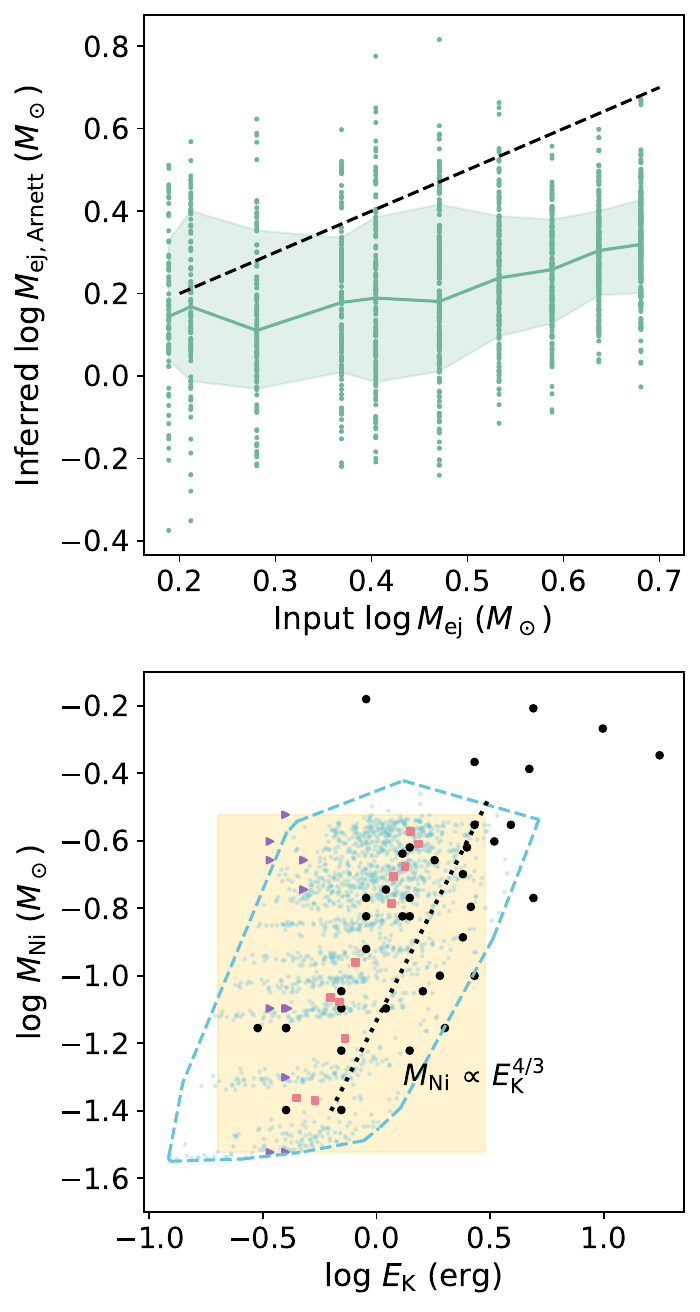}
\centering
\caption{Upper panel: Comparison between the ejecta mass $M_{\rm ej}$ of the radiation-hydrodynamic models and $M_{\rm ej}$ inferred from the Arnett-type model. The black dashed line represents $y=x$. The green solid line and the shaded region is the median value and 68\% CI, bounded by 16\% and 84\% percentiles, of the inferred $M_{\rm ej}$ after binned by the input $M_{\rm Mej}$; Lower panel: The log\,$E_{\rm K}$-log\,$M_{\rm Ni}$ space. The yellow region is the parameter space of the radiation-hydrodynamic models, and the light blue scatter points are their inferred values based on the Arnett-type model. The purple triangles represent a subset of models with similar parameters except for varied $M_{\rm Ni}$, and the pink triangles are the inferred \{$E_{\rm K}$, $M_{\rm Ni}$\} of these models. The black scatter points are SESNe data from \citet{lyman16}. The black dotted line represents $M_{\rm Ni}\propto E_{\rm K}^{4/3}$.}
\label{fig:Arnett_result}
\end{figure}

\begin{table}
\begin{center}
\caption{Parameters and priors of the Arnett-type model.}
\label{tab:priors}
\begin{tabular}{lccccc}
\hline
Parameter&Distribution&Min&Max&Mean&Std\\
\hline
$M_{\rm ej}$/$M_{\rm \odot}$ &Log flat & 0.1& 10&-&-\\ 
$v_{\rm ej}$/$10^3$\,km\,s$^{-1}$ &Gaussian & -& -&$v_{\rm Fe}$&1\\ 
$f_{\rm Ni}=\frac{M_{\rm Ni}}{M_{\rm ej}}$ &Log flat &10$^{-3}$&1&-&-\\ 
$\kappa$/cm$^2$\,g$^{-1}$ &Constant &-&-&0.06&-\\
$\kappa_{\gamma,{\rm Ni}}$/cm$^2$\,g$^{-1}$ &Log flat&10$^{-4}$&10$^{-1}$&-&-\\
$t_{\rm shift}$/day &Flat & -5& +5&-&-\\ 
\hline%
\end{tabular}
\end{center}
\end{table}

Figure~\ref{fig:Inference_Arnett_example} presents an example of an Arnett-model fit to a mock light curve generated from a radiation-hydrodynamic model. For each numerical model, we estimate the spectroscopic velocity \(v_{\rm Fe}\) as the velocity of the shell where the Sobolev optical depth of Fe~II \(\lambda5169\) is unity. This provides a model proxy for the Fe~II absorption velocity measured observationally. We use \(v_{\rm Fe}\) as the center of the Gaussian prior on \(v_{\rm ej}\) (Table~\ref{tab:priors}), following the procedure commonly adopted in observational applications of Arnett-type models. We repeat this analysis for a randomly selected subset comprising 20\% of the model grid. The comparison between the {\rm input} and inferred ejecta masses is shown in Figure~\ref{fig:Arnett_result}. We do not present the comparison for $v_{\rm ph}$, as this quantity is used to break the degeneracy between $M_{\rm ej}$ and $v_{\rm ej}$ and is therefore recovered by construction. Similarly, the comparison for $M_{\rm Ni}$ is nearly identical to that shown in Figure~\ref{fig:Scalings} and is omitted.

In this comparison, we find essentially no correlation between the inferred and the input ejecta masses, with a Spearman rank correlation coefficient of only $\rho=0.19$. This indicates that even the relative ordering of $M_{\rm ej}$ is largely lost. In practice, Arnett-model ejecta masses provide only order-of-magnitude estimates and contain little predictive power regarding the ejecta mass.

Furthermore, the inferred ejecta masses show almost no systematic dependence on the input values from the numerical models. \footnote{A recent study by \citet{yadavalli26b} reached a similar conclusion using a model grid calculated with a Monte-Carlo-based radiative-transfer code (\citealt{kasen06}). The consistency between these studies suggests that such discrepancy is not specific to the numerical models adopted here.} Instead, after binning by the input ejecta mass, the average inferred masses cluster within a relatively narrow range of $\sim1.7$-$2.5~M_\odot$, largely independent of the underlying model. Interestingly, this range is comparable to the characteristic ejecta masses inferred for SESNe in observational studies such as \cite{lyman16}. Our results therefore suggest that the apparent concentration of Arnett-model ejecta masses may arise, at least in part, from limitations of the inference framework itself rather than reflecting the intrinsic $M_{\rm ej}$ distribution of SESNe.

An additional consequence of the Arnett-model inference is a substantial distortion of the joint distribution in the $E_{\rm K}$--$M_{\rm Ni}$ plane. For the Arnett-type models, kinetic energy is estimated by assuming the ejecta is undergoing homologous expansion with constant density:
\[E_{\rm K}\,=\,\frac{3}{10}M_{\rm ej}v_{\rm ej}^2. 
\]
Here, following the observational analysis of \citet{lyman16}, the velocity spectroscopically measured from the Fe~II \(\lambda5169\), $v_{\rm Fe}$, is identified as $v_{\rm ej}$ entering the above relation. We note that this represents an approximation, as the velocity of the line-forming region does not necessarily correspond to the characteristic velocity of the global ejecta. Indeed, for our numerical models, we find \(E_{\rm K}/(M_{\rm ej}v_{\rm Fe}^{2})\,=\,0.27^{+0.05}_{-0.07}\), indicating that the coefficient is model dependent and does not exactly match the commonly adopted value of \(3/10\). Nevertheless, the purpose of this work is to evaluate the parameter inference obtained by applying the same assumptions as the observational analysis, we therefore intentionally retain the conventional coefficient of \(3/10\) to ensure a direct comparison with the results in \citet{lyman16}.

In the underlying radiation-hydrodynamic model grid, $E_{\rm K}$ and $M_{\rm Ni}$ are treated as independent parameters and therefore do not have intrinsic correlation. However, after fitting the synthetic light curves with the Arnett-type model, the inferred parameter distribution develops a pronounced sloped boundary approximately described by $M_{\rm Ni}\propto E_{\rm K}^{4/3}$, together with a morphology that differs markedly from the original distribution. This effect is illustrated in the lower panel of Figure~\ref{fig:Arnett_result}, where the light-blue region denotes the convex envelope that encompasses all the inferred pairs \{$E_{\rm K}$, $M_{\rm Ni}$\} from the Arnett-model fits, while the yellow region represents the parameter space covered by the radiation-hydrodynamic model grid.

To further demonstrate this effect, we highlight a sequence of models with similar $M_{\rm ej}$, $E_{\rm K}$ and $f_{\rm mix}$ but varying $M_{\rm Ni}$ (purple triangles in the lower panel of Figure~\ref{fig:Arnett_result}). The corresponding Arnett-model inferences are shown as pink squares. Although the original models do not contain a correlation between $E_{\rm K}$ and $M_{\rm Ni}$, a clear positive trend emerges in the inferred parameters.

The origin of this distortion can be understood from the different physical assumptions underlying the two approaches. In Arnett-type models, the diffusion timescale is assumed to be independent of $M_{\rm Ni}$ and is determined primarily by the ejecta mass and velocity. In contrast, the radiation-hydrodynamic models predict that increasing $M_{\rm Ni}$ broadens the light curve even when all other physical parameters are held fixed. This is because the additional radioactive heating modifies the ionization state of the ejecta, producing a time-dependent optical opacity $\kappa$ that is intrinsically coupled to the energy input. Consequently, the assumption in Arnett-type models of a constant opacity that is independent of the energy source breaks down. 

When the ejecta velocity is constrained observationally, the Arnett fit compensates for the broader light curve by increasing the inferred diffusion timescale. Because $v_{\rm ej}$ is effectively fixed, a longer diffusion timescale can only be achieved by increasing the inferred $M_{\rm ej}$, which in turn implies a larger inferred $E_{\rm K}$. As a result, the fitting procedure introduces a spurious correlation between $E_{\rm K}$ and $M_{\rm Ni}$. 

Interestingly, the slope of the resulting boundary in the inferred $E_{\rm K}$-$M_{\rm Ni}$ distribution closely resembles that of the observed SESNe \citep{lyman16}, where a positive correlation between kinetic energy and synthesized $^{56}$Ni mass is reported, and has often been interpreted as evidence that more energetic explosions produce larger amounts of radioactive material. Our results demonstrate that a similar morphology can emerge naturally from the inference procedure itself, even when there is no intrinsic correlation in the underlying model population. Consequently, caution is warranted when interpreting the observed $E_{\rm K}$-$M_{\rm Ni}$ relation as direct evidence for a physical connection between explosion energy and nickel production.

Overall, our results show that the dominant limitations of Arnett-type inference stem from systematic biases inherent to the model assumptions. Although the Arnett-type models provide an efficient and useful phenomenological description of SESN light curves, the inferred parameters cannot always be interpreted as direct tracers of the underlying explosion properties. Moreover, the inference procedure itself can modify parameter distributions and generate apparent correlations that are absent in the underlying models, potentially leading to misleading conclusions about the physical population. We further note that these discrepancies cannot be eliminated simply by adopting a different constant value of the optical opacity $\kappa$, one of the largest sources of uncertainty in Arnett-type modeling. Such a change would primarily introduce a systematic shift in the inferred $M_{\rm ej}$ through the \(\kappa M_{\rm ej}\) degeneracy, whereas the large model-to-model scatter and the spurious correlations shown in Figure~\ref{fig:Arnett_result} indicate that the effective opacity varies among different explosions. Capturing this behavior requires self-consistent radiation-hydrodynamic calculations. These findings highlight the need for caution when using derived parameters from the Arnett-type model to infer explosion mechanisms or progenitor evolution pathways, and the importance of physically motivated modeling frameworks for population studies of SESNe.

\section{Conclusion}
We have presented a large grid of 7,560 light-curve models for stripped-envelope supernovae (SESNe), computed with the radiation-hydrodynamic code \texttt{STELLA} based on explosions of helium-star progenitors evolved with the stellar evolution code \texttt{MESA}. The model grid spans ejecta masses motivated by nebular-phase observations, together with broad ranges of explosion energies, radioactive nickel masses, and degrees of material mixing representative of the majority of SESNe. We developed an interpolation framework for the model grid that enables rapid and continuous exploration of the parameter space while retaining the physical realism of numerical radiation-hydrodynamic simulations.

Extensive injection-recovery tests demonstrate that the interpolation is accurate, with typical uncertainties of less than $\sim0.02$ dex. We further demonstrate that the interpolation-based MCMC framework can reproduce both the bolometric light curve and photospheric velocity of the well-observed Type Ib SN 2007Y. For SN 2009jf, however, the inferred ejecta mass and degree of $^{56}$Ni mixing exhibit a strong degeneracy that cannot be fully broken even with the inclusion of color evolution. These results highlight the importance of complementary spectroscopic and nebular-phase observations for obtaining robust constraints on the physical properties of SESNe. The application of this framework to a large sample of SESNe will be presented in a forthcoming work.

We also use the numerical model grid as a controlled laboratory to evaluate the performance of the widely adopted Arnett-type analytical model. We find that the ejecta masses inferred from the Arnett-type model are not only biased, but also fail to preserve the rank ordering of the input ejecta masses across the model sample. Furthermore, the analytical model generates spurious correlations among inferred physical parameters that are absent from the underlying numerical models. These results demonstrate that, although Arnett-type models provide useful phenomenological descriptions of SESN light curves, their inferred physical parameters should be interpreted with caution, particularly in population studies where correlations among such parameters may be assigned physical significance.

Overall, this work establishes an efficient and physically motivated framework for interpreting SESN light curves with numerical radiation-hydrodynamic models. Beyond providing a practical tool for analyzing large observational samples, the framework serves as a controlled laboratory for assessing the information content of light curves and the reliability of commonly used inference techniques. As the number of well-observed SESNe continues to grow with current and forthcoming time-domain surveys, this approach provides a foundation for combining photometric and spectroscopic observations to obtain more robust constraints on the progenitors and explosion physics of stripped-envelope supernovae.

\begin{acknowledgements}
The author thank the referee for comments that helped improving the manuscript. TJM is supported by the Grants-in-Aid for Scientific Research of the Japan Society for the Promotion of Science (JP24K00682, JP21H04997, JP24H00002, JP24H00027, JP24K00668, 26H00849).

\end{acknowledgements}

\software{$\texttt{MESA}$ \citep{paxton11, paxton13, paxton15, paxton18, paxton19}; SciPy \citep{scipy}; NumPy \citep{numpy}; Astropy \citep{astropy13,astropy18}; Matplotlib \citep{matplotlib}}; $\texttt{emcee}$ \citep{emcee}; $\texttt{Redback}$ \citep{sarin24}; $\texttt{STELLA}$ \citep{blinnikov98, blinnikov00, blinnikov06}

{}
\end{document}